# The Effect of Strong Magnetic Field on Heavy Quarkonia in a hot Medium Using Nikiforov-Uvarov Method

M. Abu-shady[1] and H. M. Fath-Allah[2]

Department of Mathematics and Computer Sciences, Faculty of Science, Menoufia University, Egypt[1]

Higher Institute of Engineering and Technology, Menoufia, Egypt[2]

## Abstract

Recent analyses show that it is possible to produce a strong magnetic field at a very early stage of ultrarelativistic heavy ion collisions (URHIC), therefore, the effect of homogeneity and constant strong magnetic field on the heavy meson spectrum quarkonium is studied which the states are described as charmonium and bottominum in a non-relativistic framework and by Debye screen potential. In particular, a model that takes into account potential isotopes emerging at the level of quark-static potential, as has been observed in recent studies. Investigation is performed with and without regard to the non-isotope of fixed potential, in order to better clarify its effects.

.

## 1. Introduction

The existence of quark–gluon plasma (QGP) was explored in the mid-seventies [1, 2]. Recently, the URHIC events can be studied by combining the magnetic field effect as an extremely strong magnetic field, perpendicular to the reaction plane, is predicted to occur very early in collision phase while the event is off-central [3-7]. The magnetic field strength depends on the centrality and could be between $m_\pi^2$ ($\simeq 10^{18}$ Gauss) at RHIC [8] to $10\ m_\pi^2$ at LHC [1]. In extreme situation, it is capable of achieving values of $50\ m_\pi^2$. During the electrical weak stage transition, an incredibly large magnetic field ($\sim 10^{23}$ Gauss), was also rendered in early universe by the Higgs field gradients [9].

The disorderly measurement of heavy quarkonium potentials is important to research the dissociation of quakonia [10, 11-13]. Over the last two decades, the dynamics the dissociation of quarkonium have been seen in a medium where, in the beginning the resonance was assumed to be dissociated if the screening is strong enough, i.e. the potential is too small to keep the pair appointed by $Q\bar{Q}$. Dissociation is currently considered primarily to be due to the increase in the resonance distance either due to an inelastic mechanism of spatially mediated dispersion by a patron such as gluons, known as Landau damping [14] or because of a process glue-dissociation in which involves hard thermal gluon in one-tone color state [15]. When the medium has lower T than BE of the basic resonance, process become dominant. Thus even at lower temperature, the quarkonium is dissociated even at lower temperatures where the likelihood of color processing is small [16]. In serval approach models, the effect of $eB$ on QCD thermo-dynamics has been examined [17–38].

There are a few of studies considered the effect of magnetic field on the static characteristic of quarkonium [10, 39-45]. Heavy quarkonia has been one of the study influence on the characteristics of nuclear matter under extreme conditions, as quarkonia form in URHICs field as at a very period of $\sim 1/2m_Q$ (where $m_Q$ is the mass of the charm or bottom quark), this is equivalent to the time scale at which the magnetic field is produced. Quarkonium and heavy meson spectroscopy, with an additional spin-spin interaction term were recently mechanically investigate with vacuum quantum with harmonic oscillators and Cornell potential [39, 42] in the presence of an external magnetic field. The effect of finite T and $eB$ on the real part of the $Q\bar{Q}$ potential in the form of the destructive thermal QCD and the dissociation of heavy quarkonia due to the color screening have been studied [46]. In Ref. [16], the real part of the potential is included in the SE in order to determine the energy eigenvalues and energy eigenfunctions of the states of the $c\bar{c}$.

The aim of the present work, we have analytically solved Schrödinger equation using NU method in which the finite temperature and magnetic field are included in the potential interaction. For our best knowledge, the previous works are not solved the present potential analytically. In addition, the effect of the number of flavors is studied on the binding energy and the dissociation temperature of quarkonium.

This paper is organized as follows. In Sec. 2, we describe the NU method. In Sec.3, the method is used to solve the N-dimensional SE. In Sec. 4, we discuss the results, and summary and conclusion are given in Sec.5.

## 2. Theoretical Method

The NU method [47] used to solve the second-order differential equation in the form is defined as follows

$$\Psi''(s) + \frac{\bar{\tau}(s)}{\sigma(s)} \Psi'(s) + \frac{\breve{\sigma}(s)}{\sigma^2} \Psi(s) = 0, \tag{1}$$

where $\breve{\sigma}(s)$ and $\sigma(s)$ are max-second degree polynomials and $\bar{\tau}(s)$ is the first degree maximal polynomials. By using the transformation of $s = s(r)$,

$$\Psi(s) = \Phi(s) x(s), \tag{2}$$

as in Ref. [48], Eq. (1) can be written

$$\sigma(s) x''(s) + \tau(s) x'(s) + \lambda x(s) = 0, \tag{3}$$

where,

$$\sigma(s) = \pi(s) \frac{\Phi(s)}{\Phi'(s)}, \tag{4}$$

where, $\pi(s)$ are the first degree polynomial [47]

and

$$\tau(s) = \bar{\tau}(s) + 2 \pi(s); \quad \acute{\tau}(s) < 0, \tag{5}$$

then, the new eigenvalue equation becomes

$$\lambda = \lambda_n = -n\, \tau'(s) - \frac{n(n-1)}{2} \sigma''(s), \quad n = 0,1,2,\ldots \tag{6}$$

$x(s) = x_n(s)$ is an n degree polynomial which fulfils the form of the Rodrigues

$$x_n(s) = \frac{B_n}{\rho_n} \frac{d^n}{ds^n} (\sigma''(s) \rho(s)), \tag{7}$$

where $B_n$ is a constant of normalization and $\rho(s)$ is a function of weight that follows the next equation

$$\frac{d}{ds} \omega(s) = \frac{\tau(s)}{\sigma(s)} \omega(s), \quad \omega(s) = \sigma(s) \rho(s), \tag{8}$$

$$\pi(s) = \frac{\sigma'(s) - \bar{\tau}(s)}{2} \pm \sqrt{\left(\frac{\sigma'(s) - \bar{\tau}(s)}{2}\right)^2 - \breve{\sigma}(s) + K\,\sigma(s)}\,, \tag{9}$$

and

$$\lambda = K + \pi'(s) \tag{10}$$

$\pi(s)$ is a first degree polynomial. The $K$ in the square-root of Eq. (9) is possible to determine whether the expression under the square root is square of expression. This is possible if the discrimination is zero.

## 3. The solution of the Schrödinger radial equation in the presence of a strong magnetic field.

As in Ref. [49], in the N-dimensional space, the Schrödinger equation for two particles which interact with symmetrical potentials takes form

$$[\frac{d^2}{dr^2} + \frac{N-1}{r}\frac{d}{dr} - \frac{l(l+N-2)}{r^2} + 2\mu\,(E - V(r))]\Psi(r) = 0, \tag{11}$$

where $l$, N and $\mu$ are the angular momentum quantum number, the dimensional number, and the reduced mass. The following radial SE is obtained by applying the wave function $\Psi(r) = r^{\frac{1-N}{2}} R(r)$

$$[\frac{d^2}{dr^2} + 2\mu\,(E - V(r) - \frac{(l+\frac{N-2}{2})^2 - \frac{1}{4}}{2\mu r^2})]R(r) = 0. \tag{12}$$

In the present work, the potential takes the following form as in Ref. [50]. This potential depends on the radial distance. The effect of magnetic field will be appearing through the Debye mass. In addition, the anisotropy in the present potential with respect to the direction of magnetic field is not breaks the translational invariance of space, (see Ref. [46], for detail).

$$V(r) = \frac{-4}{3}\alpha\,(\frac{e^{-m_D\,r}}{r} + m_D) + \frac{4}{3}\frac{\sigma}{m_D}(1 - e^{-m_D r}), \tag{13}$$

where,

the string tension $\sigma = 0.18\ \text{GeV}^2$ in Ref. [16]

$$\alpha = \frac{12\,\pi}{11\,N_c\,\ln(\frac{\mu_0^2 + M_B^2}{\Lambda_V^2})} \tag{14}$$

where,

$N_c$ is the number of colors, $M_B$ (~ 1 GeV) is an infrared mass which is interpreted as the ground state mass of the two gluons bound to by the basic string, $\mu_0 = 1.1$ (GeV), $\Lambda_V = 0.385$ (GeV) as in Refs. [51-53] and the Debye mass [53] becomes

$$m_D^2 = g'^2 T^2 + \frac{g^2}{4\pi^2 T} \Sigma_f |q_f B| \int_0^\infty \frac{e^{\beta\sqrt{p_z^2+m_f^2}}}{(1+e^{\beta\sqrt{p_z^2+m_f^2}})^2} dp_z, \tag{15}$$

where,

the first term is the contribution from the gluon loops and dependent on temperature and the magnetic field doesn't affect it. The second term is this term strongly depends on the $eB$ and is not much sensitive to the T of the medium. In the first term, where $\acute{g}$ is the running strong coupling and is given by

$$\acute{g} = 4\pi \acute{\alpha}_s(T), \tag{16}$$

where,

$\acute{\alpha}_s(T)$ is the usual temperature-dependent running coupling. It is given by

$$\acute{\alpha}_s(T) = \frac{2\pi}{\left(11-\frac{2}{3}N_f\right)\ln(\frac{\Lambda}{\Lambda_{QCD}})}. \tag{17}$$

where,

$N_f$ is the number of flavors, $\Lambda$ is the renormalization scale is taken as $2\pi T$ and $\Lambda_{QCD} \sim 0.2$ (GeV) as in Ref. [16].

The second term is $g = 3.3$, $q_f$ is the quark flavor f = u and d, B is the magnetic field, $\beta$ is the inverse of temperature and quark mass massive $m_f = 0.307$ (GeV) as in Ref. [54]. In Eq. (13), $e^{-m_D r}$ is expanded up to second-order where $m_D r \ll 1$ is considered as in [55]. Eq. (13) is written as follows

$$V(r) = a_1 r^2 + a_2 r + \frac{a_3}{r} \tag{18}$$

where,

$$a_1 = \frac{-2}{3}\sigma m_D, \tag{19}$$

$$a_2 = \frac{-4}{3}\alpha m_D^2 + \frac{4}{3}\sigma, \tag{20}$$

$$a_3 = \frac{-4}{3}\alpha. \tag{21}$$

By substituting the Eq. (18) into the Eq. (12), we get

$$[\frac{d^2}{dr^2} + 2\mu(E - a_1 r^2 - a_2 r - \frac{a_3}{r} - \frac{(l+\frac{N-2}{2})^2 - \frac{1}{4}}{2\mu r^2})]R(r) = 0, \tag{22}$$

Let us assume that $r = \frac{1}{x}$ and $r_0$ is the characteristic meson radius. So we could rewrite the Eq. (22) as in Ref. [56]

$$[\frac{d^2}{dx^2} + \frac{4x}{x^2}\frac{d}{dx} + \frac{2\mu}{x^4}(E - \frac{a_1}{x^2} - \frac{a_2}{x} - a_3 x - \frac{(l+\frac{N-2}{2})^2 - \frac{1}{4}}{2\mu}x^2)]R(x) = 0, \tag{23}$$

The scheme is then based on $\frac{1}{x}$ extensions $r_0$, $y = x - \delta$ and power series around $y = 0$ where $\delta$ is a free parameter. Then, we have the scheme

$$\frac{1}{x} = \frac{1}{y+\delta} = \frac{1}{\delta}(1+\frac{y}{\delta})^{-1} \approx \frac{3}{\delta} - \frac{3x}{\delta^2} + \frac{x^2}{\delta^3},$$

$$\frac{1}{x^2} = \frac{1}{(y+\delta)^2} = \frac{1}{\delta^2}(1+\frac{y}{\delta^2})^{-2} \approx \frac{6}{\delta^2} - \frac{8x}{\delta^3} + \frac{3x^2}{\delta^4},$$

$$\tag{24}$$

Substituting Eqs. (24) into Eq. (23), we get

$$[\frac{d^2}{dx^2} + \frac{4x}{x^2}\frac{d}{dx} + \frac{2}{x^4}(-A_1 + B_1 x - C_1 x^2)]R(x) = 0, \tag{25}$$

where,

$$A_1 = -\mu(E - \frac{6 a_1}{\delta^2} - \frac{3 a_2}{\delta}), \tag{26}$$

$$B_1 = \mu(\frac{8 a_1}{\delta^3} + \frac{3 a_2}{\delta^2} - a_3), \tag{27}$$

$$C_1 = \mu(\frac{3 a_1}{\delta^4} + \frac{a_2}{\delta^3} + \frac{(l+\frac{N-2}{2})^2 - \frac{1}{4}}{2\mu}). \tag{28}$$

By comparing Eq. (25) and Eq. (1), we obtain $\bar{\tau}(s) = 4x$, $\sigma(s) = x^2$ and $\breve{\sigma}(s) = 2(-A_1 + B_1 x - C_1 x^2)$. Hence, Eq. (25) fulfils Eq. (1), therefore, the following NU method as in Sec. 2,

$$\pi = -x \pm \sqrt{-2B_1 + 2A_1(1 + K + 2 C_1)x^2} \tag{29}$$

The constant $K$ is selected as it has a double zero under the square root, i.e. its discriminate $\Delta = 4B_1 - 8 A_1(1+K + 2 C_1) = 0$. Hence,

$$\pi = -x \pm \frac{1}{\sqrt{2A_1}} (2A_1 - B_1 x). \tag{30}$$

Thus,

$$\tau = 4x \pm 2(-x + \frac{1}{\sqrt{2A_1}} (2A_1 - B_1 x)). \tag{31}$$

In the above equation, we select the positive sign to have a derivative

$$\tau' = 2 - \frac{2B_1}{\sqrt{2A_1}} \tag{32}$$

using the Eq. (10), we get

$$\lambda = \frac{B_1^2}{2A_1} - 2C_1 - \frac{B_1}{\sqrt{2A_1}} - 2 \tag{33}$$

and Eq. (6), we obtain

$$\lambda_n = -n\left(2 - \frac{2B_1}{\sqrt{2A_1}}\right) - n(n-1) \tag{34}$$

From Eq. (6), $\lambda = \lambda_n$.

$$\frac{B_1^2}{2A_1} - 2C_1 - \frac{B_1}{\sqrt{2A_1}} - 2 = -n\left(2 - \frac{2B_1}{\sqrt{2A_1}}\right) - n(n-1) \tag{35}$$

Let, $z = \frac{B_1}{\sqrt{2A_1}}$ then the equation becomes quadratic

$$z^2 - (2n+1)z + (n^2 + n - 2(C_1 + 1)) = 0 \tag{36}$$

The Eq. (36) is solved, we get the spectrum of energy

$$E = \frac{6a_1}{\delta^2} + \frac{3a_2}{\delta} - \frac{2\mu(\frac{8a_1}{\delta^3} + \frac{3a_2}{\delta^2} - a_3)^2}{\left[(2n+1) \pm \sqrt{1 + 8\mu \left(\frac{3a_1}{\delta^4} + \frac{a_2}{\delta^3} + \frac{(l+\frac{N-2}{2})^2 - \frac{1}{4}}{2\mu}\right)}\right]^2} \tag{37}$$

## 4. Results and Discussion

We note that Debye screening depends on eB and T. In Fig. (1), we find $m_D$ increases with the temperature linearly and the effect the $N_f$ plays a role in increasing the Debye mass with temperature. In the left panel, when $m_f > T$ and $T^2 < eB$, the Debye mass increases with temperature. In right panel, when $m_f < T$ and $T^2 < eB$, the Debye mass increases with temperature and we note that the quark mass does not affect on the Debye mass. This conclusion is in an agreement with Ref. [57].

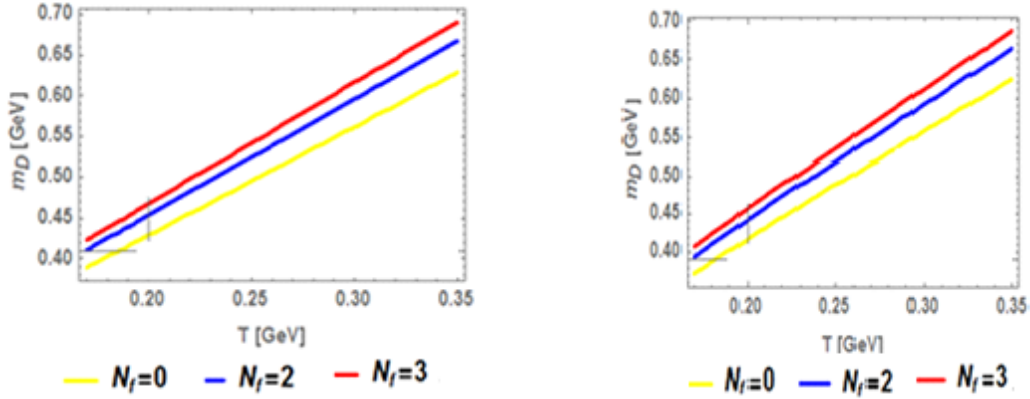

**Fig. (1).** *Left panel: the Debye mass is plotted with T for different values the number of flavors at a fixed value of B($eB = 10\ m_\pi^2$) and quark mass ($m_f = 0.307\ GeV$). Right panel: the Debye mass is plotted for the different values of the number of flavors at a fixed value of B ($eB = 10\ m_\pi^2$) and quark masses ($m_f = 0.025(GeV)$).*

In Fig. (2), we note that $m_D$ increases with T and increases the number of flavors. In left panel, when $m_f > T$, $T^2 < eB$, the Debye mass increases with increasing temperature. In the right panel, when $m_f < T$ and $T^2 < eB$, the Debye mass increases with increases to temperature. Therefore, we noted that increasing magnetic field up to $15\ m_\pi^2$ is not affected on the behavior of the Debye mass. Also, this conclusion is an agreement with Ref. [16].

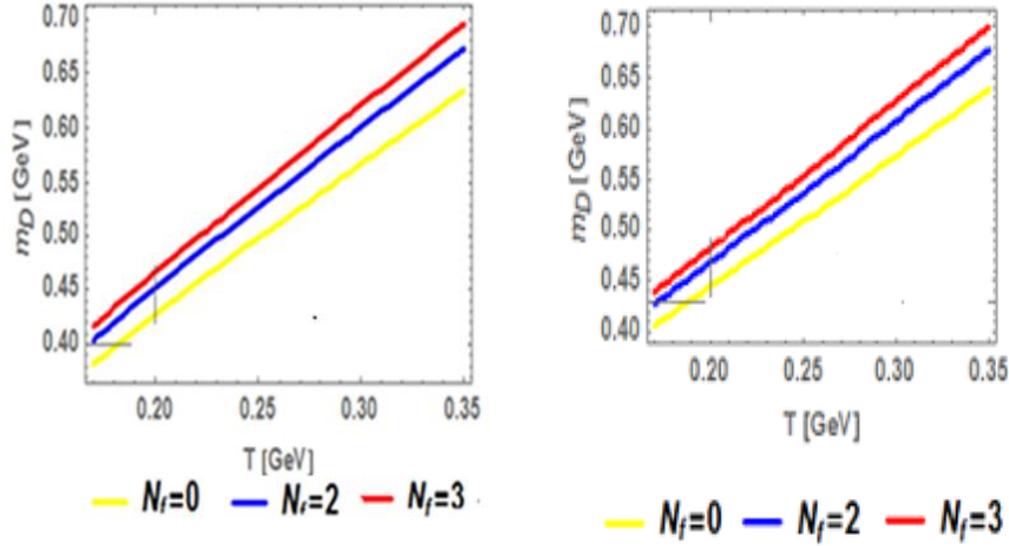

**Fig. (2).** *Left panel: the screening Debye mass is plotted as a function of T for different the number of flavors at a fixed value of B($eB = 15\, m_\pi^2$) and quark masses ($m_f = 0.307(GeV)$). Right panel: the Debye screening mass with T for different values of the number of flavors at a fixed value of B $eB = 15\, m_\pi^2$) and quark masses ($m_f = 0.025(GeV)$).*

In Fig. (3), in the left panel the $m_D$ increases with increasing T and $eB$ with ($N_F = 2$) and quark masses massive ($m_f = 0.307(GeV)$). We show that the $m_D$ increases linearly both with T and the $eB$. We took quark masses mass ($m_f = 0.025(GeV)$). In the right panel, we have plotted the Debye screening mass with temperature and magnetic field B with ($N_F = 2$), we find that the quark mass does not affect on the Debye mass.

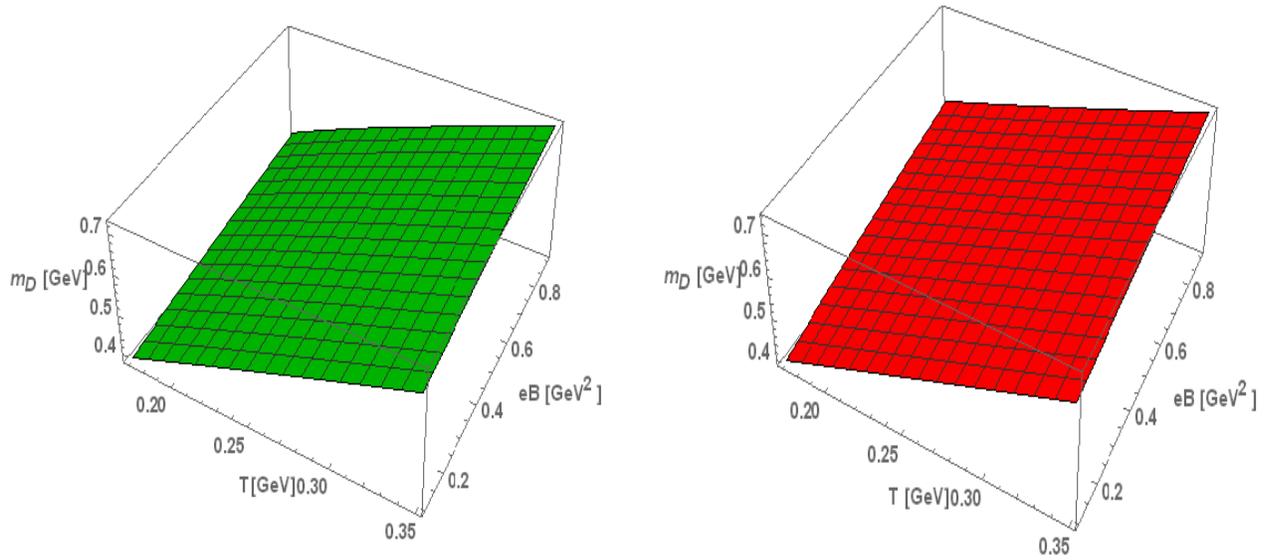

**Fig. (3).** *In the left panel the Debye screening mass is plotted with temperature and magnetic field B at $(N_F = 2)$ and quark masses $(m_f = 0.307(GeV))$. In the right panel, the Debye screening mass is plotted with temperature and magnetic field B at $(N_F = 2)$ and quark masses $(m_f = 0.025(GeV))$.*

In Fig.(4), we have plotted the Debye screening mass with temperature and magnetic field B with $(N_F = 0, N_F = 2)$ and quark masses $(m_f = 0.307(GeV))$. We show that the $m_D$ increases linearly with both T and $eB$. By increasing the number of flavors, the Debye mass increases with temperature and magnetic field.

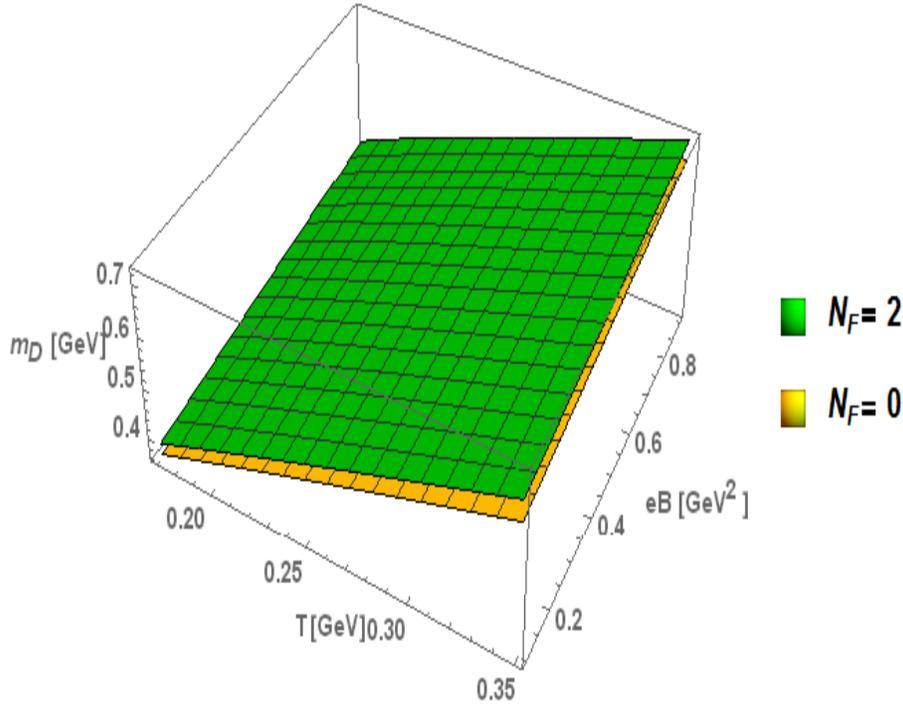

**Fig. (4).** *The Debye screening mass is plotted with temperature and magnetic field B at ($N_F = 0$, $N_F = 2$) and quark masses ($m_f = 0.307 (GeV)$).*

Also, in this section, we note the behavior of the real potential that plays a fundamental role in the current work. The quark-antiquark interaction potential is plotted a function of distance (r), where the T and $eB$ is included in the potential through the $m_D$. The Debye screening mass is parameterized according to Eq. (15) in which $N_c$=3, $N_f = 2$ and g = 3.3 are taken.

In Fig. (5), in the left panel, we plotted the real part of the potential as a function of r for different values of the magnetic field like $eB = 10\ m_\pi^2$, $eB = 25 m_\pi^2$ and $eB = 50 m_\pi^2\ at$ the fixed value of temperature $T = T_c$. We noted that when the value of the $eB$ is increased the real-part is more screened. We noted that the ($eB = 10 m_\pi^2$) has an effect on the linear term. However, a further increase of magnetic field ($eB = 25 m_\pi^2$) and ($eB = 50 m_\pi^2$) the potential becomes more attractive than $eB = 10 m_\pi^2$. In the right panel, we have plotted the real part of the potential as a function of different temperature values like T= 1.5 $T_c$, T= 2 $T_c$, and T= 3 $T_c$ for a fixed value of $eB = 10\ m_\pi^2$. We have seen that the potential is screened by increasing the temperature. As a result, the real part of the potential was found to be more screened to increase the value of both T and $eB$. This conclusion has been agreed with Ref. [16].

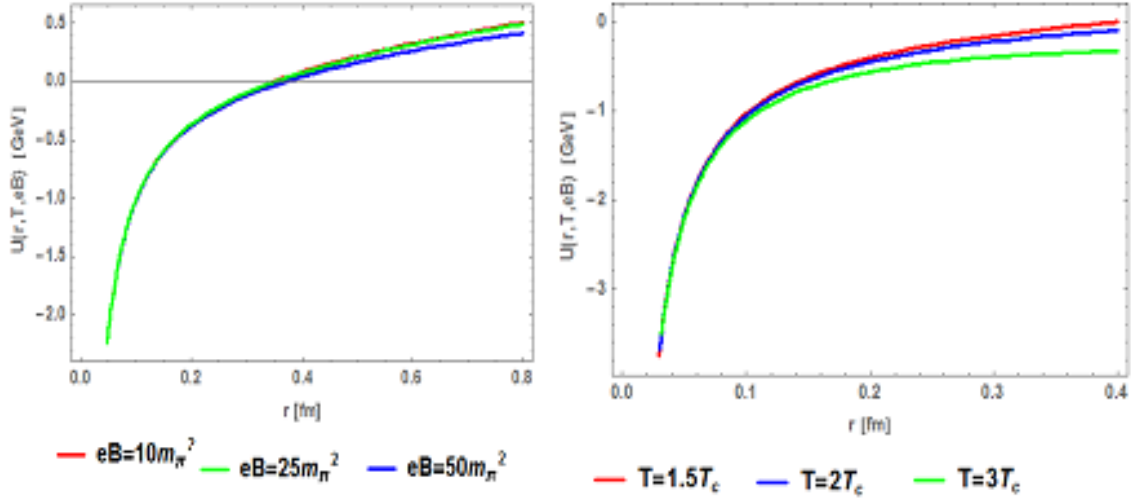

**Fig. (5).** *Potential interaction is plotted as a distance function (r) for the different magnetic field values. In the right panel, the potential interaction is plotted as a distance function (r) for different potential temperature values.*

In Fig. (6), we have plotted the real part of potential as a function of the temperature and the magnetic field for the fixed value of r = 0.2 fm. To see the change of the potential with the strong magnetic field for the temperature range T= 0.17- 0.3 GeV, we notice that potential is more attractive with a magnetic field than temperature. This conclusion is in agreement with [58].

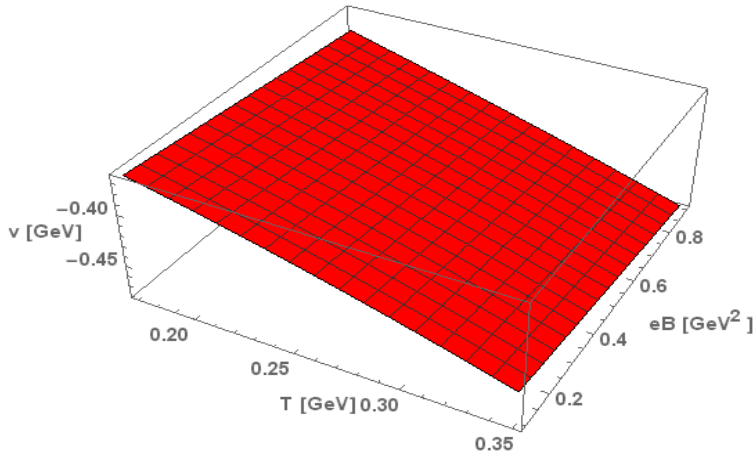

**Fig. (6).** *Effect the temperature and magnetic field on the potential.*

## 4.1. Binding energy

By solving Schrodinger equation as discussed in Sec. 3. We need to mention that the radial Schrödinger equation is numerically solved as in Refs. [46, 50]. We obtain the BEs of $c\bar{c}$ and $b\bar{b}$. In following, we see the change of the binding energy under the effect of temperature and magnetic field.

Charmonium binding energy is plotted as a function of T for three cases $eB = 5\,m_\pi^2$, $eB = 25 m_\pi^2$ and $eB = 50 m_\pi^2$. In Fig. (7), we show the effect of the $eB$, temperature, and number of flavors on the BE of charmonium. We find that the BEs decreases with increasing T and magnetic field decrease. Besides, we have seen that the effect of temperature is more effective than the extremal magnetic field. This conclusion is in an agreement with works [16, 50]. Also, the binding energy decreases with increasing the number of flavors ($N_f$) as shown in Fig. (7). The effect is not considered in other works.

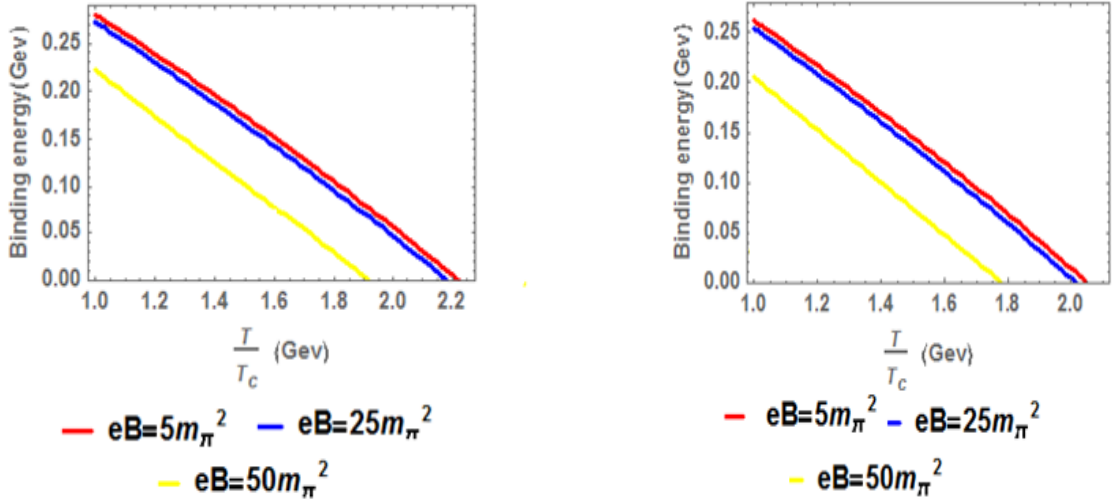

**Fig. (7).** *In the left panel, the Binding energy of charmonium is plotted as a function of the T in the thermal medium in the presence of the eB for the different magnetic field values at $N_f = 0$. In the right panel, the Binding energy of charmonium is plotted as a function of the T in the thermal medium in the presence of the eB for the different magnetic field values at $N_f = 2$.*

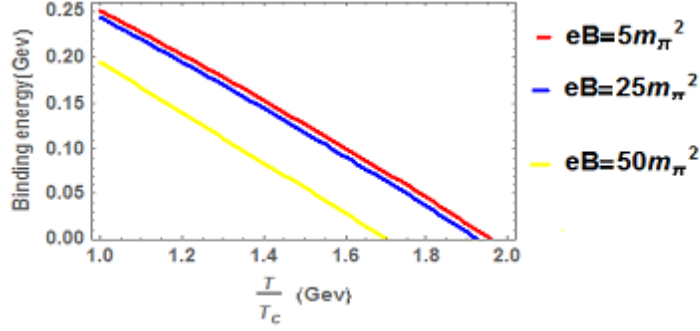

*In the lower panel, the Binding energy of charmonium is plotted as a function of T in the thermal medium in the presence of the eB for the different magnetic field values at $N_f = 3$.*

In Fig.(8), we have plotted the binding energy of charmonium at temperature T= 1.98 $T_c$ ,T= 1.99 $T_c$, and T= 2 $T_c$ as a function of the magnetic field. We find that BE decreases with the magnetic field increases. By increasing the temperature, we notice the binding energy decreases. This conclusion is in agreement with works [16, 58].

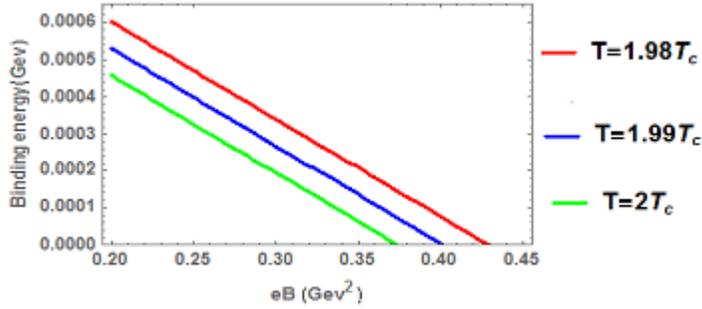

**Fig. (8).** *The binding energy of charmonium is plotted as a function of the magnetic field in the thermal medium for different values of the temperature at $N_f = 2$.*

In Fig.(9), bottmonium binding energy is plotted as a function of T for three cases $eB = 5\, m_\pi^2$ ,$eB = 25 m_\pi^2$ and $eB = 50 m_\pi^2$. By increasing the magnetic field, we notice that the binding energy of 1S bottomonium decreases. In the upper left panel, we took $N_f = 0$. Besides, the binding energy decreases when taking $N_f = 2$ in the upper right panel. Finally, we find that the BEs decrease with the increase of $N_f$. At $N_f = 3$, as shown in the lower panel. As a result, we deduce that the $N_f$ plays a role in the decrease of the BE. This finding is in agreement with works

[16, 50]. A similar situation also observed for charmonium, except that the BE for charmonium is higher than that for bottomonium.

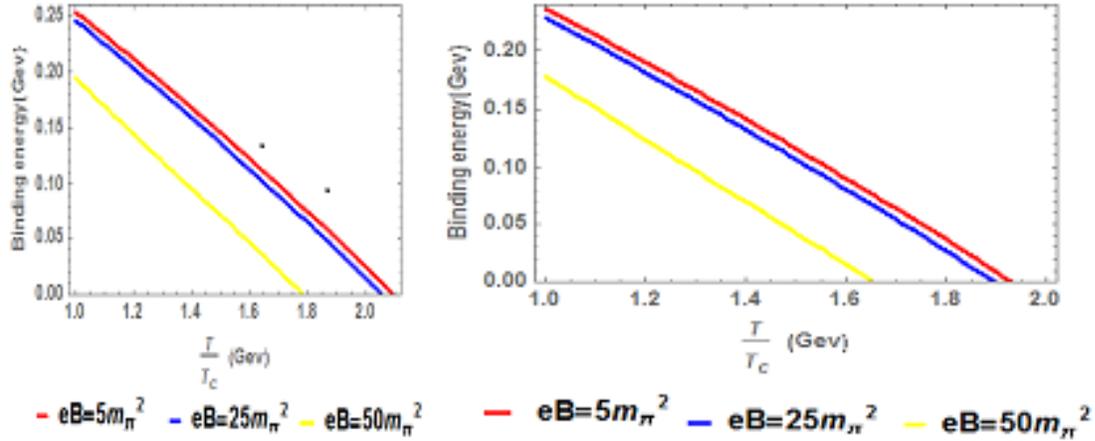

**Fig. (9).** *In the upper left panel, the Binding energy of bottomonium is plotted as a function of the temperature in the thermal medium in the presence of magnetic field for the different magnetic field values at $N_f = 0$. In the upper right panel, the binding energy of bottomonium (in GeV) is plotted as a function of the temperature in the thermal medium in the presence of magnetic field for the different magnetic field values at $N_f = 2$.*

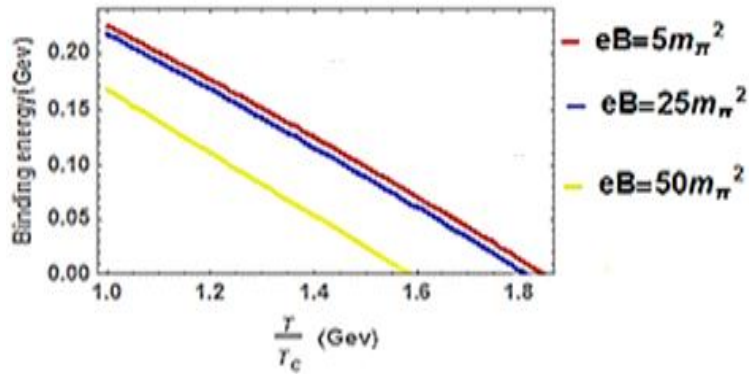

*In the lower panel, bottomonium Binding energy (in GeV) is plotted as a function of the temperature in the thermal medium in the presence of magnetic field for the different magnetic field values at $N_f = 3$.*

In Fig.(10), we have plotted the binding energy of bottomonium at temperature T= 1.98 $T_c$, T= 1.99 $T_c$, and T= 2 $T_c$ as a function of the $eB$. Note that BE with the $eB$ decreases. By increasing the temperature, we notice the binding energy decreases. Also, we note that the binding energy tends to zero that depends on the value of the temperature of the medium. This result is noted in Ref. [ 16,50].

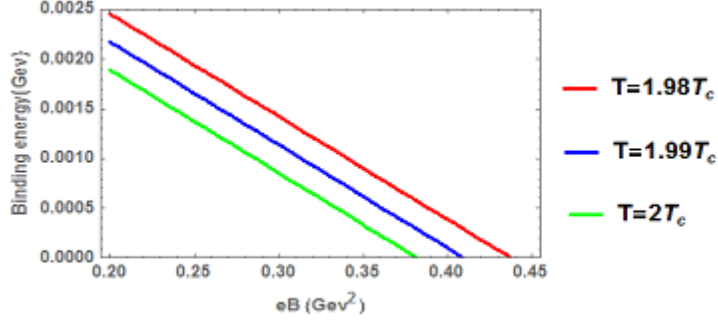

**Fig. (10).** *The binding energy of bottomonium (in GeV) is plotted as a function of the magnetic field in the thermal medium for different values of the temperature at $N_f$ =2.*

## 4.2. Dissociation temperature for heavy quarkonia

The dissociation of the binding state of the two bodies in a thermal medium has improved over the past two decades, as it was initially thought that the resonance would separate when the scan became strong enough, i.e. the potential is too small to carry $Q\bar{Q}$ pair together. The dissociation is currently considered primarily to be due to the increase in the resonance distance either due to an inelastic mechanism of spatially mediated dispersion by a patron such as gluons, known as Landau damping [14] or because of a process glue-dissociation in which involves hard thermal gluon in one-tone color state [15]. When the medium has lower T than BE of the basic resonance, process become dominant. Thus even at lower temperature, the quarkonium is dissociated even at lower temperatures where the likelihood of color processing is small.

In the present study, we calculate the dissociation temperature at $E_b \simeq 0$, an approximation that provides good accuracy in the calculation of the dissociation temperature. In the current analysis, we also study influence of the $eB$ on the dissociation temperature in the presence of hot media for charmonium and bottomonium, using the calculated binding energies.

In Table (1), the dissociation temperature ($T_D$) is the minimum when the magnetic field at $eB = 50\ m_\pi^2$. When $N_f = 0$, the charmonium is dissociated at 2.22 $T_c$ when $eB = 5\ m_\pi^2$. At the case, $eB = 25\ m_\pi^2$ and $eB = 50\ m_\pi^2$, are dissociated at 2.19 $T_c$ and 1.91 $T_c$. When $N_f = 2$ charmonium is dissociated at 2.05 $T_c$ when $eB = 5\ m_\pi^2$, in the case, $eB = 25\ m_\pi^2$ and $eB = 50\ m_\pi^2$, are dissociated at 2.01 $T_c$ and 1.77 $T_c$. We notice that $T_D$, in this case is lower compared to $N_f = 0$. When $N_f = 3$ charmonium is dissociated at 1.96 $T_c$ when $eB = 5\ m_\pi^2$, in the case, $eB = 25\ m_\pi^2$ and $eB = 50\ m_\pi^2$ are dissociated at 1.93 $T_c$ and 1.7 $T_c$.

In the Table (2), at $N_f = 0$ bottomonium is dissociated with 2.1 $T_c$ when $eB = 5\ m_\pi^2$, in the case, $eB = 25\ m_\pi^2$ and $eB = 50\ m_\pi^2$, are dissociated at 2.05 $T_c$ and 1.79 $T_c$. When $N_f = 2$ bottomonium is dissociated at 1.94 $T_c$ when $eB = 5\ m_\pi^2$, in the case, $eB = 25\ m_\pi^2$ and $eB = 50\ m_\pi^2$, are dissociated at 1.9 $T_c$ and 1.65 $T_c$. We note that $T_D$ is lower compared to $N_f = 0$. When $N_f = 3$ bottomonium is dissociated at 1.85 $T_c$ when $eB = 5\ m_\pi^2$, in the case, $eB = 25\ m_\pi^2$ and $eB = 50\ m_\pi^2$, are dissociated at 1.81 $T_c$ and 1.59 $T_c$. This conclusion is agreed with Ref. [16, 50]. In Ref. [50], In the SE, it is used to the real part of potential and they have found that the real part of the potential is more screened and by increasing in the screening of the real part of the potential leads to the decrease of BEs of Y and J/Ψ. Finally, they got the $T_D$ for Y and J/Ψ, which became slightly lower in the presence of a weak magnetic field. At eB= $0.5 m_\pi^2$ they dissociated at slightly lower value 1.13$T_c$ and 3.94$T_c$. In Ref. [16], the $eB$ influences the binding of J/ψ and $\chi_c$. it reduces the binding of J/ψ but increases the binding of $\chi_c$. In the other hand, the magnetic field raises the width of the resonances because the temperature is too high. We eventually obtained the dissociation due to the Landau damping and noted that the $T_D$ had risen in the presence of a $eB$. At eB = $6m_\pi^2$ the J/ψ is dissociated at $2 T_c$, and with eB = $4m^2\pi$, the $\chi_c$ is dissociated at $1.1 T_c$. As the $eB$ increases further the $T_D$ decreases.

**Table. 1.** Dissociation temperature ($T_D$) for charmonium.

| State | $eB = 5\ m_\pi^2$ | $eB = 25\ m_\pi^2$ | $eB = 50\ m_\pi^2$ |
|---|---|---|---|
| $N_f = 0$ | $2.22\ T_c$ | $2.19\ T_c$ | $1.91\ T_c$ |
| $N_f = 2$ | $2.05\ T_c$ | $2.01\ T_c$ | $1.77\ T_c$ |
| $N_f = 3$ | $1.96\ T_c$ | $1.93\ T_c$ | $1.7\ T_c$ |

**Table. 2.** Dissociation temperature ($T_D$) for bottomonium.

| State | $eB = 5\ m_\pi^2$ | $eB = 25\ m_\pi^2$ | $eB = 50\ m_\pi^2$ |
|---|---|---|---|
| $N_f = 0$ | $2.1\ T_c$ | $2.05\ T_c$ | $1.79\ T_c$ |
| $N_f = 2$ | $1.94\ T_c$ | $1.9\ T_c$ | $1.65\ T_c$ |
| $N_f = 3$ | $1.85\ T_c$ | $1.81\ T_c$ | $1.59\ T_c$ |

### 4.3. Dissociation of heavy quarkonia in a magnetic field

We calculate the dissociation of charmonium and bottomonium in the magnetic field when $E_b \simeq 0$.

**Table. 3.** Dissociation of charmonium in the magnetic field.

| $c\bar{c}$ | $T = 2T_c$ | $T = 1.99\ T_c$ | $T = 1.98T_c$ |
|---|---|---|---|
|  | $eB = 19.4\ m_\pi^2$ | $eB = 21\ m_\pi^2$ | $eB = 22.57\ m_\pi^2$ |

By taking thermal medium at $T = 2T_c$, we note that the binding energy dissociated as magnetic field increases $eB = 19.4\ m_\pi^2$. By decreasing the temperature of the medium up to $T = 1.98\ T_c$, we note that the binding energy dissociated at eB = $22.57\ m_\pi^2$. The state of bottomonium is similar to that states of Table (4) but the dissociation of bottomonium is more than of charmonium. This conclusion is agreed with works such that [15, 59].

**Table. 4.** Dissociation of bottomonium in the magnetic field.

| $b\bar{b}$ | $T = 2\, T_c$ | $T = 1.99\, T_c$ | $T = 1.98\, T_c$ |
|---|---|---|---|
| | $eB = 19.95\, m_\pi{}^2$ | $eB = 21.5\, m_\pi{}^2$ | $eB = 23\, m_\pi{}^2$ |

## 5. Summary and conclusion

In this paper, we find the dissociation of heavy quarks in hot QCD plasma in the presence of a strong magnetic field using a generalized $m_D$ that depends on T and $eB$. The SE is solved analytically using NU method, where the real potential includes the influence of the finite T and $eB$.

We consider the effect of the number of flavors, finite temperature, and magnetic field on binding energy and dissociation temperature. We found that the $N_f$ has a basic role on decreasing binding energy. We have observed that the magnetic field is largely affected by large-distance interaction, as a result of which the real part of potential is more attractive. We report on the results for the values of the magnetic field at $eB = 5\, m_\pi{}^2$, $eB = 25 m_\pi{}^2$ and $eB = 50 m_\pi{}^2$. We found the binding energy decreases by increasing the magnetic field. Also, we measure that the $T_D$ is above the critical temperature $T_c = 0.17$ GeV, and that the $T_D$ of charmonium and bottomonium is lower in a strong magnetic field. This occurs because of the BE decreases by increasing the magnetic field. We note that the $T_D$ decreases with an increase in the $N_f$ and decrease with magnetic field values. We note that the dissociation temperature the of charmonium is greater than that of the bottomonium since the mass of charmonium is smaller than the mass of the bottomonium. This conclusion is agreed with results of Refs. [16, 46, 50, 59].

**Data Availability:** The data used to support the findings of this study are included within the article and are cited at relevant places within the text as references.

**Conflicts of Interest:** The author declares that he has no conflicts of interest.